**Research status and trends in Operations Research and Management Science (OR/MS) journals: A bibliometric analysis based on the Web of Science database 2001-2012**

Pablo Dorta-González * [a], María Isabel Dorta-González [b], Dolores Rosa Santos-Peñate [c], Rafael Suárez-Vega [d]

[a] Universidad de Las Palmas de Gran Canaria, TiDES Research Institute, Facultad de Economía, Empresa y Turismo, Campus de Tafira, 35017 Las Palmas de Gran Canaria, Spain. *E-mail*: pdorta@dmc.ulpgc.es

[b] Universidad de La Laguna, Departamento de Ingeniería Informática, Avenida Astrofísico Francisco Sánchez s/n, 38271 La Laguna, Spain. *E-mail*: isadorta@ull.es

[c] Universidad de Las Palmas de Gran Canaria, TiDES Research Institute, Facultad de Economía, Empresa y Turismo, Campus de Tafira, 35017 Las Palmas de Gran Canaria, Spain. *E-mail*: drsantos@dmc.ulpgc.es

[d] Universidad de Las Palmas de Gran Canaria, TiDES Research Institute, Facultad de Economía, Empresa y Turismo, Campus de Tafira, 35017 Las Palmas de Gran Canaria, Spain. *E-mail*: rsuarez@dmc.ulpgc.es

* Corresponding author. E-mail: pdorta@dmc.ulpgc.es (P. Dorta-González).

## Abstract

A bibliometric analysis to evaluate global scientific production in the subject category of Operations Research and Management Science (OR/MS) from 2001 to 2012 was applied. Data was based on the Web of Science (Science Citation Index) database compiled by Thomson Reuters. The results showed that the OR/MS research has significantly increased over the past twelve years. The Bradford core journals in the category were identified. The researchers paid great attention to "networks", "control", and "simulation". Among the countries, USA attained a dominant position in global research in the field.

## Introduction

The subjects *Operations Research* (OR) and *Management Science* (MS) (even though there may be philosophical differences, we use the two terms interchangeably), have been defined by many authors in the field. Definitions range from "a scientific approach to decision making" to "the use of quantitative tools for systems originated from real life situations" (Eiselt & Sandblom, 2010). The Institute for Operations Research and Management Sciences (INFORMS) defines OR/MS as the science of the optimum solutions. What all of this essentially means is that the science indeed uses quantitative techniques to make and prepare decisions, by determining the most efficient way to act under given circumstances. In other words, rather than throwing large amounts of resources (such as money) at a problem, OR/MS will determine ways to do things more efficiently.

As a formal discipline, OR/MS originated from the efforts by military planners during World War II. In the decades after the war, the techniques began to be applied more widely to problems in business, industry, and society. Since that time, OR/MS has expanded into a wide range of industries ranging from petrochemicals to airlines, finances, logistics, and governments, focusing on the development of mathematical models that can be used to analyse and optimize complex systems, and has become an area of active academic and industrial research (Eiselt & Sandblom, 2010).

For decades, the *journal impact factor* (IF) has been an accepted indicator in ranking journals, however, there are increasing arguments against the fairness of using the IF as the sole ranking criteria (Waltman & Van Eck, 2013). The *2-year impact factor* published by Thomson Reuters in the *Journal Citation Reports* (JCR) is defined as the average number of references to each journal in a current year with respect to 'citable items' published in that journal during the two preceding years (Garfield, 1972).

Since its formulation, the IF has been criticized for some arbitrary decisions involved in its construction. The definition of 'citable items' (these include letters and peer reviewed papers –articles, proceedings papers, and reviews–), the focus on the two preceding years as representation of impact at the research front, etc., have been discussed in the literature (Bensman, 2007; Moed et al., 2012) and have given rise to suggestions of many possible modifications and improvements (Bornmann & Daniel, 2008; Dorta-González & Dorta-González, 2013a,b,c; Dorta-González et al., 2014).

Journal performance is a complex multi-dimensional concept difficult to be fully captured in one single metric (Moed et al., 2012, p. 368). This has resulted in the creation of many other quality metric indices, such as the fractionally counted impact factor (Leydesdorff & Bornmann, 2011), the audience factor (Zitt & Small, 2008), the source normalized impact per paper (Moed, 2010), the scimago journal ranking (González-Pereira et al., 2009), the h-index (Hirsch, 2005), and the central area index (Dorta-González & Dorta-González, 2011; Egghe, 2013) to name a few.

Some bibliometric studies have been carried out on particular aspects of OR/MS. In the literature, Reisman & Kirschnick (1994) survey the contents of three journals in the US, Operations Research, Management Science, and Interfaces, to conclude the direction that OR/MS should take. In a later paper, Ormerod & Kiossis (1997) extend the analysis to the UK. The journals analysed include Journal of the Operational Research Society, Omega, OR Insight, and European Journal of Operational Research. Eto (2000, 2002) compares the contents and citation patterns of the Journal of the Operational Research Society, Operations Research, and Management Science. Chang & Hsieh (2008) survey the OR/MS research in Asia. Kao (2009) investigates 56 OR/MS journals from all over the world, with emphasis on the contribution of different countries to OR/MS studies and the country spread of each journal. Finally, White et al. (2011) survey the OR/MS research in the developing countries.

However these studies were mainly based on an analysis of a part, meanwhile there is no report which indicates the research status and trends of the academic field as a whole. In this paper we investigate the scientific progress in the discipline. We analyse the subject category OR/MS to illustrate global research status and trends in the discipline during the period 2001–2012. The result could help to better understand the global development of the OR/MS discipline, and potentially guide scientists towards evaluating and orienting their research.

## Materials and methodology

The underlying bibliometric data was obtained from the online version of the *Web of Science* database (Science Citation Index) during the last week of February 2013. This database (reported by Thomson Reuters – ISI, Philadelphia, USA) is available at the website www.webofknowledge.com. All journals (77) listed in the subject category of Operations Research and Management Science were considered. The impact factors (IF) of the journals were obtained from the year 2011 Journal Citation Reports (JCR), which were the latest data available.

With the intention of measuring and evaluate the scientific progress at the research front, we have focused on the research articles which are the primary source of the research results. All the articles referring to the OR/MS subject category during 2001–2012 were assessed taking into account the following aspects: document type and language, countries, institutions, journal, author keywords, citations, and authors h-index. The h-index is a simple measure incorporating both quantity and quality that has many advantages over other bibliometric measures.

Collaboration type was determined by the addresses of the authors, where the term "single country publication" was assigned if the researchers' addresses were from the same country. The term "international collaborative publication" was designated to those papers that were coauthored by researchers from more than one country. Similarly, the term "single institution publication" was assigned if the researchers' addresses were from the same institution. The term "inter-institutional collaborative publication" was assigned if authors were from different institutions.

NetDraw (Borgatti, 2002) is a free program and one of the best network visualization packages on the market. It was used to analyse the international collaborations and inter-institutional collaborations.

## Results and discussion

### *Document type and language*

Following the Web of Science's document type classification, nine document types were found in a total of 71,670 publications during 2001–2012, as displayed in Fig. 1. Articles (66,283) were the most-frequently used document type comprising 92.5% of the total production, followed by proceedings papers (4,874, 6.8%), editorial materials (2,558, 3.6%), and reviews (868, 1.2%). The others were less significant and below 1%, including book reviews (713), corrections (396), news items (358), letters (325), and biographical items (128). From the point of language analysis, most of papers were published in English, comprising 99.9% of the total publications.

[Fig. 1 about here]

*Distribution by output*

In total, there were 66,283 research articles downloaded from the Web of Science database during 2001–2012 in the subject category of Operations Research and Management Science. The number of articles consistently increased in this period, as displayed in Fig. 2, with just an exception in year 2010. A total of 3,824 articles were published in 2001, while the number of articles was 7,972 in 2012, about twice that in 2001. Note the fast-growing period 2004–2009 in which the growth rate reached 18%.

[Fig. 2 about here]

*Distribution by journal output*

A total of 66,283 research articles were published in 77 scientific journals under the subject category OR/MS from 2001 to 2012. Comparison of major factors (i.e. total publications, number of papers in 2012, two-years and five-years impact factors, and h-index) among all journals is listed in Table 1. Among these journals, the IF and h-index are highly variable. Expert Systems with Applications published the majority of papers not only those in the past twelve years (6,936, 10.5%) but also in 2012, while Journal of Operations Management had the highest h-index (102) and IF (4.382 and 6.012).

[Table 1 about here]

Regarding impact factor, note that the optimum citation time window in the category OR/MS is greater than two years because in the majority of cases the five-years impact factor is greater than the two-years impact factor.

In addition, Bradford's Law of Scattering (Bradford, 1934) was applied. Bradford's law is a pattern first described by Samuel C. Bradford in 1934 that estimates the exponentially diminishing returns of extending a search for references in science journals. The journals were sorted in descending order in terms of number of papers, and then divided into three approximately equal parts. The first part containing 5 (6.5%) out of 77 journals represents the most productive one-third of the total papers. The second part with 16 (20.8%) out of 77 journals represents the moderate productive one-third of total papers, and the third part with 56 (72.7%) out of 77 journals representing the least productive one-third of total papers. The number of journals in the three zones approximately followed Bradford's law and was close to $1{:}n{:}n^2$ (1:3.3:10.9). Expert Systems with Applications, European Journal of Operational Research, International Journal of Production Research, International Journal of Production Economics, and Journal of the Operational Research Society were the five Bradford's core journals in the OR/MS subject category.

*Distribution by countries/territories*

The top 20 most productive countries/territories are shown in Table 2. Among the top 20 countries/territories were nine Asian countries, seven European countries, three American countries, and one Oceanic country. The USA ranked first in terms of: total, single country and international collaborative publications. However, the collaborative papers represented only 33.6% of the total publications from the USA, which was less than most other countries in the top 20. Canada ranked 4th in the number of total publications, but it had the highest proportion (57.6%) of collaborative papers. The publication impact of the USA was excellent with the highest h-index (114) among all the countries, followed by China (66) and Canada (62). In addition, half of the countries had an index of 36–46. Notice that Taiwan ranked 3th in the number of total publications, but had a lower h-index of 43. However, Australia ranked 14th in the number of total publications, but had a higher h-index of 49.

[Table 2 about here]

Fig. 3, produced from NetDraw program (Borgatti, 2002), shows the collaboration network of the top 50 countries/territories based on the number of research articles. NetDraw measures the relative importance of nodes within the network and could be viewed as an indicator of countries' positions within the collaboration network in our study. The thickness of links shows the strength of correlation, and the size of nodes shows the amount of single country publication. In this study, a core group of countries in the collaboration network is visualized. The United States took the central position in the network of international collaboration, and it was the principal collaborator with major productive countries, such as China and Canada (Fig. 3).

[Fig. 3 about here]

*Distribution by author keywords*

The author keywords of total articles in the Web of Science database during 2001–2012 were considered. The top 25 most frequently used author keywords and their rank and percentage in different periods are listed in Table 3, which shows the mainstream research in OR/MS is mainly focused on “networks”, “control”, “simulation”, “production”, and so on.

[Table 3 about here]

Notice “genetic algorithms” only ranked 20th in the 2001–2004 period, but it ranked 11th in the 2009–2012 period, which indicates that this issue might be one of the research hot topics in the coming years. Similarly, the rank of “scenarios”, “logistics”, and “supply chain management” raised from 27th, 31th, 32th in 2001–2004 period to 18th, 19th, 22th in 2009–2012 period, respectively.

However, “manufacturing” ranked 3th in 2001–2004, only ranked 8th in the 2009–2012 period. This indicates that this issue might be a frozen research topic in the coming years. Something similar occur with “queueing” which ranked 11th in the 2001–2004 period, but it ranked 24th in the 2009–2012 period.

A keywords tag crowd with the top 100 words is shown in Fig. 4. This was elaborated with a free on-line application available in TagCrowd.com.

[Fig. 4 about here]

*Most frequently cited papers*

The top 20 most cited papers (articles and reviews) are listed in Table 4. Among all the papers referring to the subject category of OR/MS in the period 2001–2012, there were 527,737 total citations and an average of 7.96 citations per paper. However, only 28.6% of all papers were above average. The most frequently cited paper was “On the implementation of an interior-point filter line-search algorithm for large-scale nonlinear programming”, which was published in Mathematical Programming by Wachter, A, and Biegler, LT (2006) from IBM Corp and Carnegie Mellon University, respectively, and cited 456 times by 2012. The second most frequently cited paper titled “Variable neighborhood search: Principles and applications” was published in the European Journal of Operational Research in 2001. It was written by Hansen, P, and Mladenovic, N, and cited 438 times by 2012.

[Table 4 about here]

*Most productive authors*

The top 20 most productive authors are listed in Table 5. Among the top 20 authors, ten were from China, followed respectively by the USA (3), Canada (2), Israel (1), Taiwan (1), India (1), South Korea (1), and Singapore (1). Cheng T.C.E who published the greatest number of papers had an h-index of 25, and was ranked 1st. Levitin ranked 5th in the number of papers, he had only an h-index of 15, and was ranked 11th. Yao who ranked 10th in the number of papers had an h-index of 19, and was ranked 3th.

[Table 5 about here]

## Conclusions

For the bibliometric analysis in the subject category of Operations Research and Management Science, we obtained some significant results on the global research performance throughout the period from 2001 to 2012. The number of research articles consistently increased in this period. Especially noteworthy was 2004–2009 in which the growth rate reached 18%. Expert Systems with Applications published the most papers not only in the past twelve years but also in 2012, while the Journal of Operations Management had the highest h-index and IF. The citation maturity time in the OR/MS category is greater than two years and the production follows the Bradford’s law with five journals in its core.

Among the top 20 countries/territories nine were in Asia. The USA ranked first in terms of total, single country and international collaborative publications. However, the collaborative papers represented only 33.6% of the total publications from the USA, which was less than most other countries in the top 20. The publication impact of the USA was excellent with the highest h-index (114) among all countries, followed by China (66) and Canada (62). Among the top 20 most productive authors, ten were from China, five from North America, and the rest from Asian countries.

The mainstream research in OR/MS is mainly focused on “networks”, “control”, “simulation”, and “production”, although “genetic algorithms” might be one of the research hot topics in the coming years.

Finally, we have tried to measure the scientific progress in the discipline. These results could help to better understand the global development of the discipline, and potentially guide scientists towards evaluating and orienting their research.

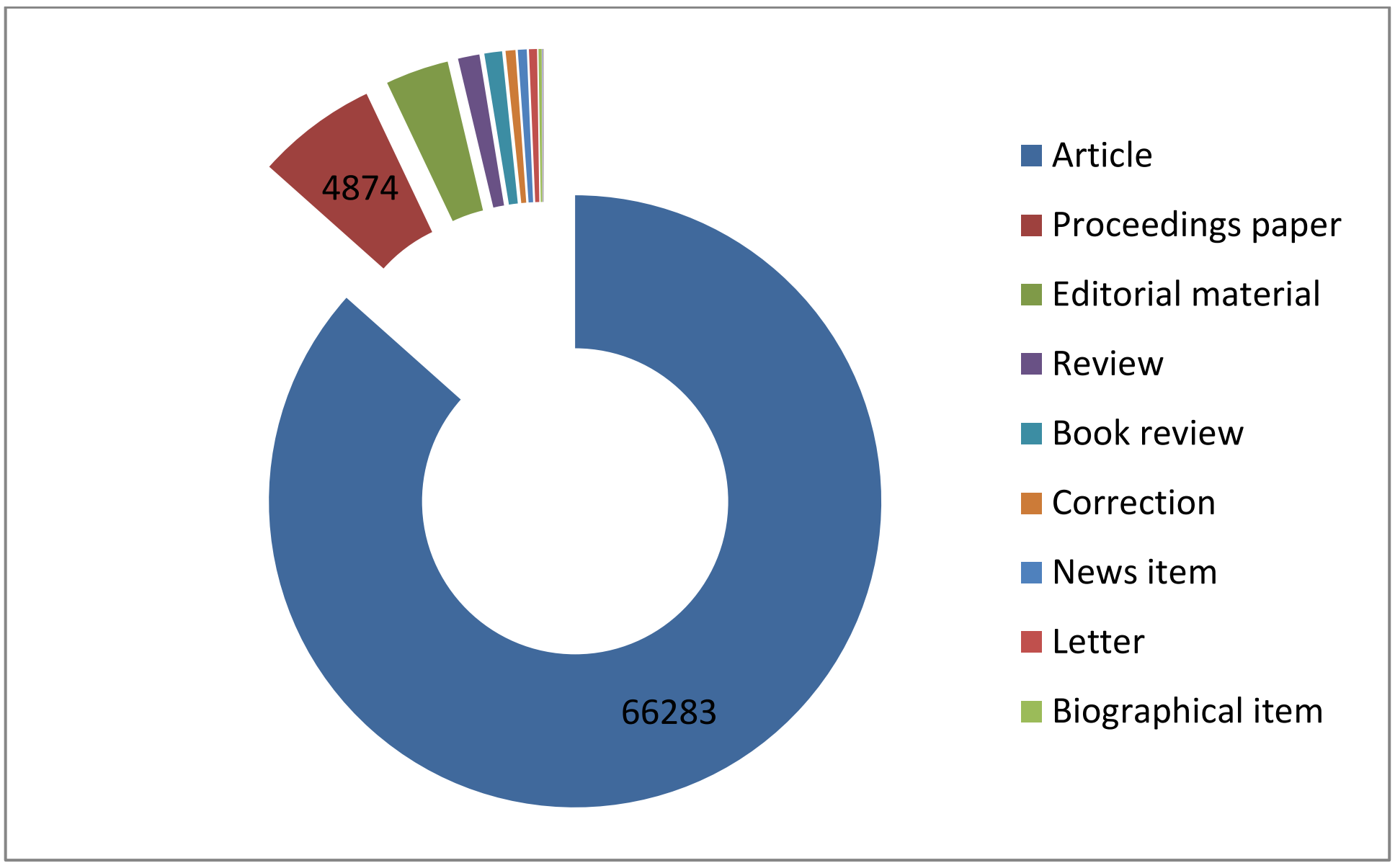

**Fig. 1** Document types in journals under the SCI category of OR/MS from 2001 to 2012

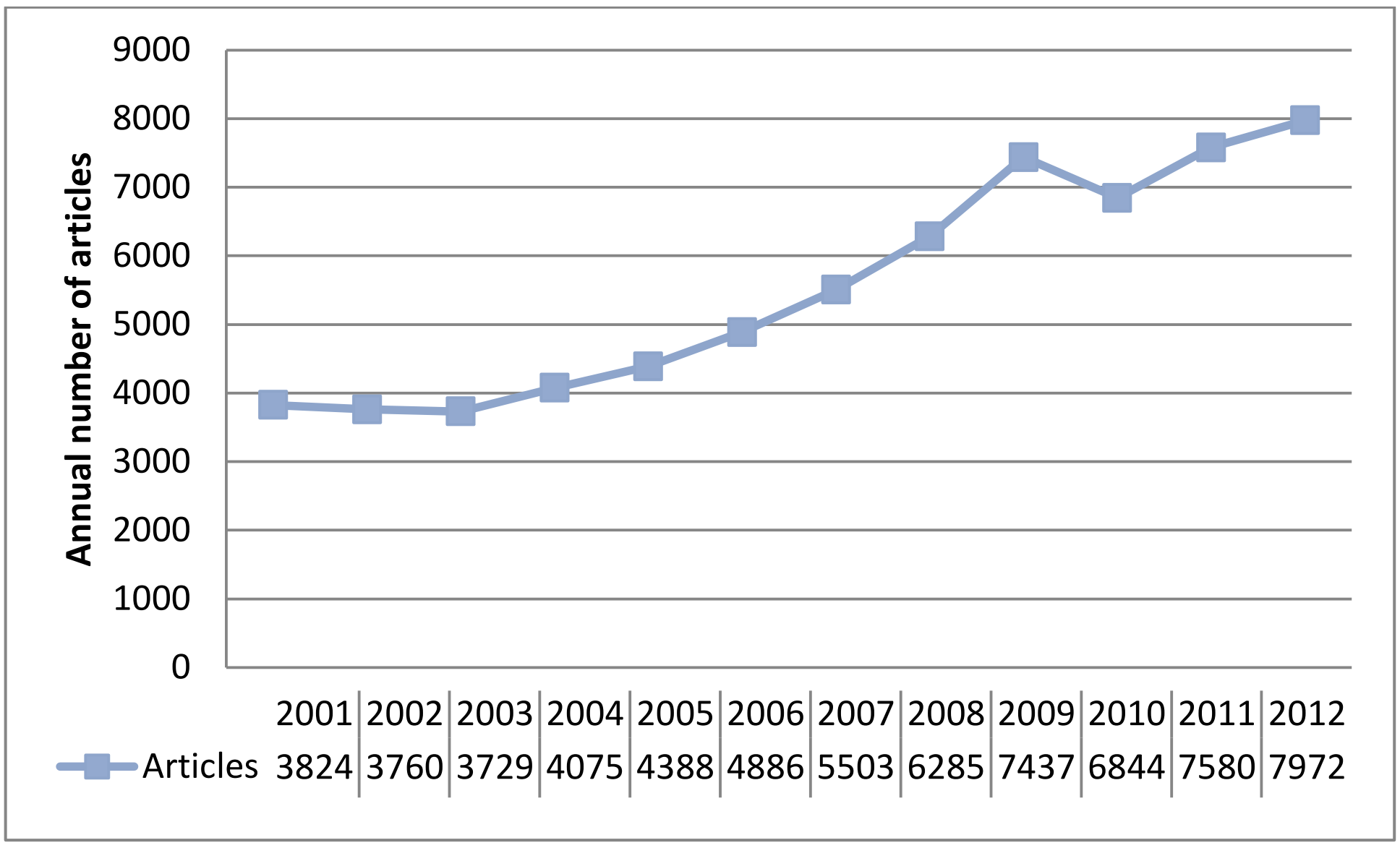

**Fig. 2** Annual distribution of research articles in journals under the SCI category of OR/MS from 2001 to 2012

**Table 1** Comparison among all journals in the SCI category of OR/MS during 2001-2012

| *Journal* | *TP (%)* | *TP 2012* | *IF-2 (rank)* | *IF-5* | *h-index (rank)* |
|---|---|---|---|---|---|
| Expert Syst Appl | 6,936 (10.46) | 1,709 | 2.203 (5) | 2.455 | 83 (5) |
| Eur J Oper Res | 6,913 (10.43) | 455 | 1.815 (6) | 2.277 | 96 (2) |
| Int J Prod Res | 3,732 (5.63) | 365 | 1.115 (25) | 1.367 | 83 (5) |
| Int J Prod Econ | 2,872 (4.33) | 303 | 1.760 (8) | 2.384 | 89 (3) |
| J Oper Res Soc | 2,408 (3.63) | 192 | 0.971 (35) | 1.350 | 67 (11) |
| Comput Oper Res | 2,333 (3.52) | 172 | 1.720 (10) | 1.984 | 54 (16) |
| Reliab Eng Syst Safe | 1,804 (2.72) | 163 | 1.770 (7) | 2.170 | 64 (13) |
| J Optimiz Theory App | 1,726 (2.60) | 145 | 1.062 (28) | 1.200 | 45 (28) |
| Ann Oper Res | 1,631 (2.46) | 153 | 0.840 (41) | 1.101 | 36 (50) |
| Manage Sci | 1,630 (2.46) | 136 | 1.733 (9) | 3.304 | 77 (8) |
| Syst Control Lett | 1,459 (2.20) | 136 | 1.222 (23) | 1.718 | 38 (45) |
| Int J Syst Sci | 1,419 (2.14) | 178 | 0.991 (33) | 1.257 | 47 (22) |
| Decis Support Syst | 1,391 (2.10) | 148 | 1.687 (12) | 2.331 | 46 (26) |
| J Global Optim | 1,275 (1.92) | 130 | 1.196 (24) | 1.391 | 49 (20) |
| Safety Sci | 1,261 (1.90) | 159 | 1.402 (19) | 1.578 | 69 (9) |
| Oper Res | 1,153 (1.74) | 116 | 1.665 (13) | 2.285 | 55 (15) |
| Oper Res Lett | 1,146 (1.73) | 92 | 0.537 (55) | 0.821 | 29 (63) |
| Int J Technol Manage | 1,125 (1.70) | 74 | 0.516 (57) | 0.702 | 78 (7) |
| Math Program | 1,043 (1.57) | 78 | 1.707 (11) | 2.182 | 54 (16) |
| IIE Trans | 1,027 (1.55) | 63 | 0.856 (39) | 1.469 | 47 (22) |
| Interfaces | 988 (1.49) | 38 | 0.843 (40) | 1.048 | 39 (40) |
| Technovation | 979 (1.48) | 56 | 3.287 (3) | 2.760 | 86 (4) |
| J Syst Eng Electron | 955 (1.44) | 142 | 0.276 (73) | N/A | 30 (61) |
| Qual Reliab Eng Int | 868 (1.31) | 96 | 0.700 (45) | 0.842 | 47 (22) |
| Prod Plan Control | 854 (1.29) | 54 | 0.725 (43) | 0.841 | 61 (14) |
| Comput Optim Appl | 774 (1.17) | 82 | 1.350 (20) | 1.432 | 29 (63) |
| Omega Int J Manage S | 763 (1.15) | 74 | 3.338 (2) | 3.622 | 68 (10) |
| Transport Res B Meth | 746 (1.13) | 98 | 2.856 (4) | 3.393 | 54 (16) |
| Optimization | 702 (1.06) | 92 | 0.500 (58) | 0.677 | 39 (40) |
| Nav Res Log | 681 (1.03) | 56 | 1.038 (31) | 1.278 | 42 (34) |
| Math Method Oper Res | 672 (1.01) | 42 | 0.481 (61) | 0.684 | 37 (47) |
| Networks | 662 (1.00) | 64 | 0.983 (34) | 1.022 | 39 (40) |
| Eng Optimiz | 648 (0.98) | 69 | 0.936 (37) | 1.077 | 26 (68) |
| Transport Res E Log | 645 (0.97) | 85 | 1.648 (14) | 2.126 | 54 (16) |
| Queueing Syst | 630 (0.95) | 59 | 0.717 (44) | 0.983 | 41 (36) |
| Optim Method Softw | 613 (0.92) | 47 | 0.651 (48) | 0.744 | 45 (28) |
| Int J Comput Integ M | 601 (0.91) | 82 | 1.071 (27) | 1.113 | 49 (20) |
| Math Oper Res | 593 (0.89) | 42 | 1.056 (29) | 1.398 | 44 (32) |
| Prod Oper Manag | 572 (0.86) | 63 | 1.301 (21) | 2.259 | 66 (12) |
| J Oper Manag | 543 (0.82) | 53 | 4.382 (1) | 6.012 | 102 (1) |
| Informs J Comput | 512 (0.77) | 46 | 1.076 (26) | 1.260 | 45 (28) |
| Appl Stoch Model Bus | 506 (0.76) | 49 | 0.690 (46) | 0.736 | 15 (72) |
| J Ind Manag Optim | 438 (0.66) | 61 | 0.661 (47) | 0.749 | 37 (47) |
| Transport Sci | 422 (0.64) | 35 | 1.507 (16) | 2.107 | 46 (26) |
| Asia Pac J Oper Res | 421 (0.64) | 44 | 0.250 (75) | 0.427 | 14 (73) |
| Probab Eng Inform Sc | 415 (0.63) | 31 | 0.642 (51) | 0.627 | 33 (57) |
| J Qual Technol | 407 (0.61) | 24 | 1.564 (15) | 1.860 | 36 (50) |
| OR Spectrum | 381 (0.57) | 38 | 1.233 (22) | 1.706 | 45 (28) |
| Optim Lett | 368 (0.56) | 56 | 0.952 (36) | 0.908 | 32 (59) |
| J Manuf Syst | 346 (0.52) | 24 | 0.639 (53) | 0.928 | 40 (37) |
| Concurrent Eng Res A | 323 (0.49) | 25 | 0.478 (62) | 0.710 | 17 (71) |
| Optim Contr Appl Met | 318 (0.48) | 43 | 0.648 (49) | 0.891 | 34 (54) |
| IEEE Syst J | 309 (0.47) | 55 | 0.923 (38) | 0.916 | 42 (34) |
| Optim Eng | 304 (0.46) | 34 | 0.476 (63) | 1.008 | 39 (40) |
| Infor | 273 (0.41) | 11 | 0.295 (72) | 0.596 | 40 (37) |
| M&Som Manuf Serv Op | 268 (0.40) | 35 | 1.475 (18) | 2.356 | 47 (22) |
| Discrete Optim | 257 (0.39) | 44 | 0.500 (58) | 0.696 | 10 (74) |
| Top | 250 (0.38) | 24 | 0.765 (42) | 1.067 | 34 (54) |
| Rairo Oper Res | 249 (0.38) | 12 | 0.220 (77) | 0.278 | 34 (54) |
| Netw Spat Econ | 241 (0.36) | 34 | 1.019 (32) | 1.658 | 40 (37) |
| Discrete Event Dyn S | 230 (0.35) | 19 | 0.641 (52) | 0.979 | 20 (70) |
| Pac J Optim | 214 (0.32) | 37 | 0.527 (56) | N/A | 33 (57) |
| Stud Inform Control | 211 (0.32) | 39 | 0.578 (54) | N/A | 30 (61) |
| J Syst Sci Syst Eng | 209 (0.32) | 28 | 0.429 (64) | N/A | 39 (40) |
| Mil Oper Res | 193 (0.29) | 17 | 0.343 (70) | 0.314 | 29 (63) |
| Q J Oper Res | 185 (0.28) | 32 | 0.323 (71) | N/A | 10 (74) |
| Int T Oper Res | 178 (0.27) | 38 | 0.648 (49) | N/A | 38 (45) |
| J Scheduling | 171 (0.26) | 44 | 1.051 (30) | 1.497 | 36 (50) |
| Cent Eur J Oper Res | 157 (0.24) | 36 | 0.484 (60) | N/A | 6 (76) |
| Eur J Ind Eng | 154 (0.23) | 24 | 0.413 (67) | 0.415 | 37 (47) |
| P I Mech Eng O J Ris | 154 (0.23) | 37 | 0.393 (68) | N/A | 35 (53) |
| Fuzzy Optim Decis Ma | 141 (0.21) | 21 | 1.488 (17) | N/A | 31 (60) |
| Qual Technol Quant M | 124 (0.19) | 31 | 0.276 (73) | N/A | 28 (67) |
| Systems Eng | 123 (0.19) | 30 | 0.420 (66) | N/A | 43 (33) |
| Flex Serv Manuf J | 71 (0.11) | 18 | 0.250 (75) | 0.450 | 29 (63) |
| Stat Oper Res T | 69 (0.10) | 15 | 0.429 (64) | N/A | 26 (68) |
| Eng Econ | 61 (0.09) | 17 | 0.382 (69) | N/A | 3 (77) |

*TP (%)* total number and percentage of publications (articles), *IF* impact factor in 2011 (two-years and five-years citation time windows)

**Table 2** Top 20 most productive countries/territories with articles in journals under the SCI category of OR/MS during 2001–2012

| *Country /territory* | *TP* | *TPR (%)* | *SP* | *SPR (%)* | *CP* | *CPR (%)* | *RC (%)* | *h-index* |
|---|---|---|---|---|---|---|---|---|
| USA | 19,021 | 1 (28.70) | 12,621 | 1 (19.04) | 6,400 | 1 (9.66) | 15 (33.6) | 114 |
| China | 8,776 | 2 (13.24) | 5,596 | 2 (8.44) | 3,180 | 2 (4.80) | 11 (36.2) | 66 |
| Taiwan | 5,241 | 3 (7.91) | 4,391 | 3 (6.62) | 850 | 11 (1.28) | 20 (16.2) | 43 |
| Canada | 4,044 | 4 (6.10) | 1,716 | 8 (2.59) | 2,328 | 3 (3.51) | 1 (57.6) | 62 |
| England | 4,018 | 5 (6.06) | 1,982 | 5 (2.99) | 2,036 | 4 (3.07) | 5 (50.7) | 56 |
| France | 3,469 | 6 (5.23) | 1,773 | 7 (2.67) | 1,696 | 5 (2.56) | 7 (48.9) | 60 |
| Spain | 3,088 | 7 (4.66) | 2,068 | 4 (3.12) | 1,020 | 8 (1.54) | 17 (33.0) | 52 |
| Germany | 2,991 | 8 (4.51) | 1,710 | 9 (2.58) | 1,281 | 6 (1.93) | 9 (42.8) | 53 |
| Italy | 2,867 | 9 (4.33) | 1,871 | 6 (2.82) | 996 | 9 (1.50) | 13 (34.7) | 46 |
| Netherlands | 2,386 | 10 (3.60) | 1,217 | 12 (1.84) | 1,169 | 7 (1.76) | 6 (49.00) | 46 |
| South Korea | 2,301 | 11 (3.47) | 1,532 | 10 (2.31) | 769 | 12 (1.16) | 16 (33.4) | 42 |
| Turkey | 1,929 | 12 (2.91) | 1,266 | 11 (1.91) | 663 | 14 (1.00) | 14 (34.4) | 40 |
| Japan | 1,844 | 13 (2.78) | 1,195 | 13 (1.80) | 649 | 15 (0.98) | 12 (35.2) | 36 |
| Australia | 1,751 | 14 (2.64) | 807 | 15 (1.22) | 944 | 10 (1.42) | 4 (53.9) | 49 |
| India | 1,741 | 15 (2.63) | 1,168 | 14 (1.76) | 573 | 17 (0.86) | 18 (32.9) | 37 |
| Singapore | 1,268 | 16 (1.91) | 572 | 19 (0.86) | 696 | 13 (1.05) | 3 (54.9) | 40 |
| Israel | 1,173 | 17 (1.77) | 613 | 18 (0.92) | 560 | 18 (0.84) | 8 (47.7) | 44 |
| Belgium | 1,162 | 18 (1.75) | 524 | 20 (0.79) | 638 | 16 (0.96) | 2 (54.9) | 42 |
| Brazil | 1,110 | 19 (1.67) | 695 | 17 (1.05) | 415 | 19 (0.63) | 10 (37.4) | 32 |
| Iran | 1,074 | 20 (1.62) | 778 | 16 (1.17) | 296 | 20 (0.45) | 19 (27.6) | 24 |

*TP* total publications, *SP* single country publications, *CP* international collaborative publications, *R* rank, *RC* rank in percentage of international collaborative publications in total publications of each country

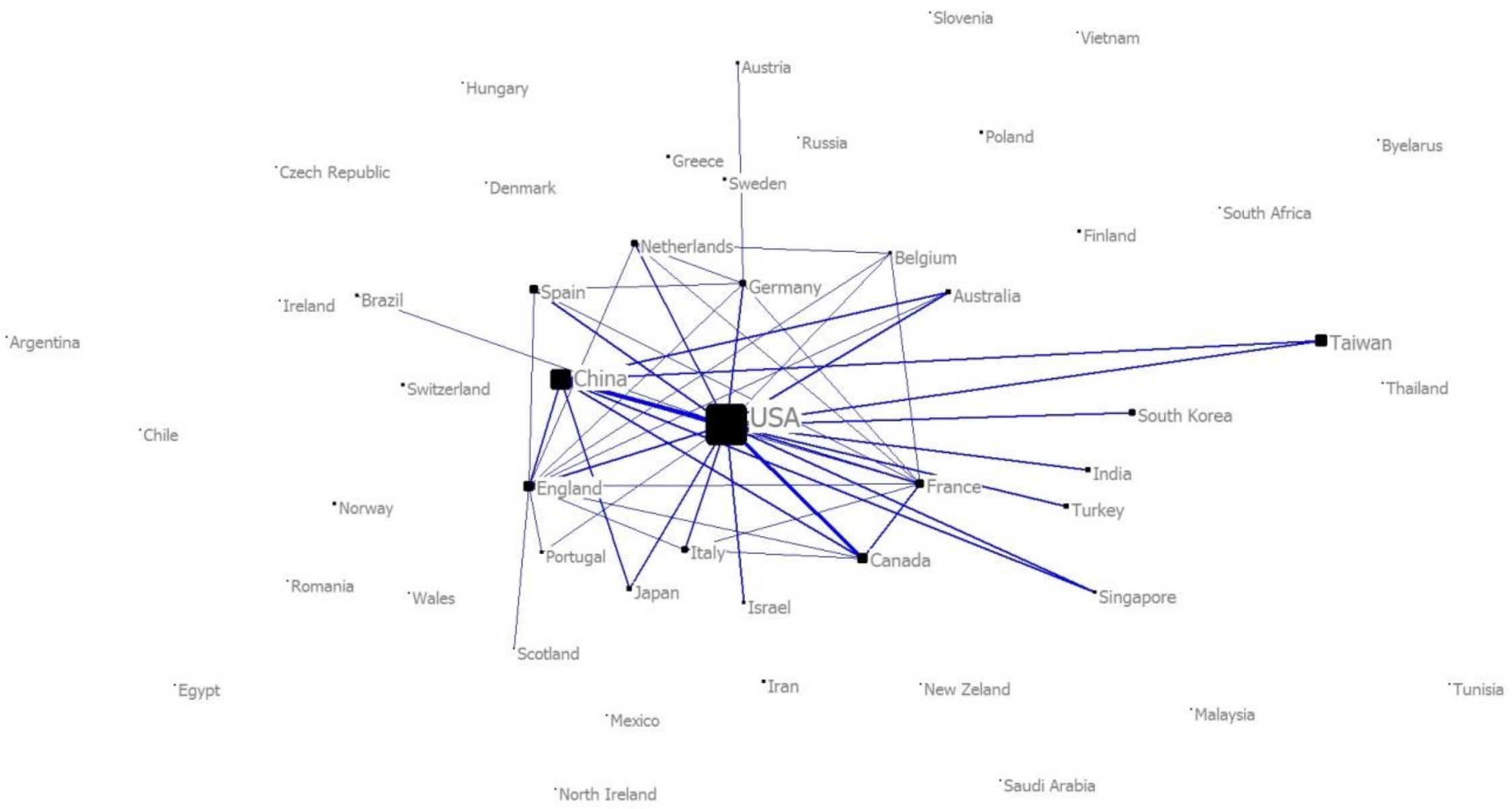

**Fig. 3** The network of international collaborations (80 or more) among the top 50 countries/territories based on the total number of research articles published in journals under the SCI category of OR/MS from 2001 to 2012

**Table 3** Top 25 most frequent author keywords used in articles published in all journals under the SCI category of OR/MS during 2001–2012

| *Keyword* | *2001–2012* | *2001–2012* *R (%)* | *2001–2004* *R (%)* | *2005–2008* *R (%)* | *2009–2012* *R (%)* |
|---|---|---|---|---|---|
| Networks | 11,890 | 1 (7.16) | 2 (5.87) | 1 (6.66) | 1 (8.11) |
| Control | 10,255 | 2 (6.17) | 1 (6.89) | 2 (5.95) | 2 (5.98) |
| Simulation | 7,166 | 3 (4.31) | 8 (3.95) | 8 (4.07) | 3 (4.65) |
| Production | 7,026 | 4 (4.23) | 4 (4.94) | 4 (4.27) | 5 (3.86) |
| Scheduling | 6,967 | 5 (4.19) | 5 (4.62) | 3 (4.41) | 6 (3.84) |
| Distribution | 6,823 | 6 (4.11) | 6 (4.32) | 6 (4.12) | 4 (4.00) |
| Manufacturing | 6,486 | 7 (3.90) | 3 (5.04) | 7 (4.08) | 8 (3.24) |
| Heuristics | 6,451 | 8 (3.88) | 7 (4.01) | 5 (4.15) | 7 (3.64) |
| Inventory | 4,997 | 9 (3.01) | 9 (3.04) | 9 (3.26) | 10 (2.82) |
| Pricing | 4,630 | 10 (2.79) | 12 (2.23) | 10 (2.90) | 9 (2.97) |
| Linear programming | 3,968 | 11 (2.39) | 10 (2.75) | 11 (2.42) | 13 (2.19) |
| Integer programming | 3,753 | 12 (2.26) | 14 (2.13) | 13 (2.33) | 12 (2.27) |
| Constraint programming | 3,711 | 13 (2.23) | 13 (2.20) | 12 (2.39) | 14 (2.14) |
| Genetic algorithms | 3,541 | 14 (2.13) | 20 (1.50) | 16 (1.89) | 11 (2.60) |
| Location | 3,359 | 15 (2.02) | 15 (1.97) | 14 (1.97) | 15 (2.08) |
| Routing | 3,057 | 16 (1.84) | 17 (1.71) | 15 (1.91) | 16 (1.85) |
| Reliability | 2,937 | 17 (1.77) | 16 (1.82) | 17 (1.72) | 17 (1.78) |
| Queueing | 2,744 | 18 (1.65) | 11 (2.26) | 18 (1.68) | 24 (1.35) |
| Assignment | 2,598 | 19 (1.56) | 18 (1.64) | 19 (1.62) | 20 (1.49) |
| Transportation | 2,449 | 20 (1.47) | 22 (1.39) | 20 (1.51) | 21 (1.48) |
| Scenarios | 2,287 | 21 (1.38) | 27 (1.05) | 23 (1.27) | 18 (1.61) |
| Cutting | 2,239 | 22 (1.35) | 19 (1.52) | 21 (1.38) | 25 (1.24) |
| Logistics | 2,200 | 23 (1.32) | 31 (0.95) | 25 (1.19) | 19 (1.59) |
| Supply chain management | 2,091 | 24 (1.26) | 32 (0.95) | 22 (1.30) | 22 (1.38) |
| Decision analysis | 2,085 | 25 (1.25) | 24 (1.13) | 26 (1.19) | 23 (1.36) |

*TP* total publications, *R (%)* rank and percentage of the author keywords in total publications

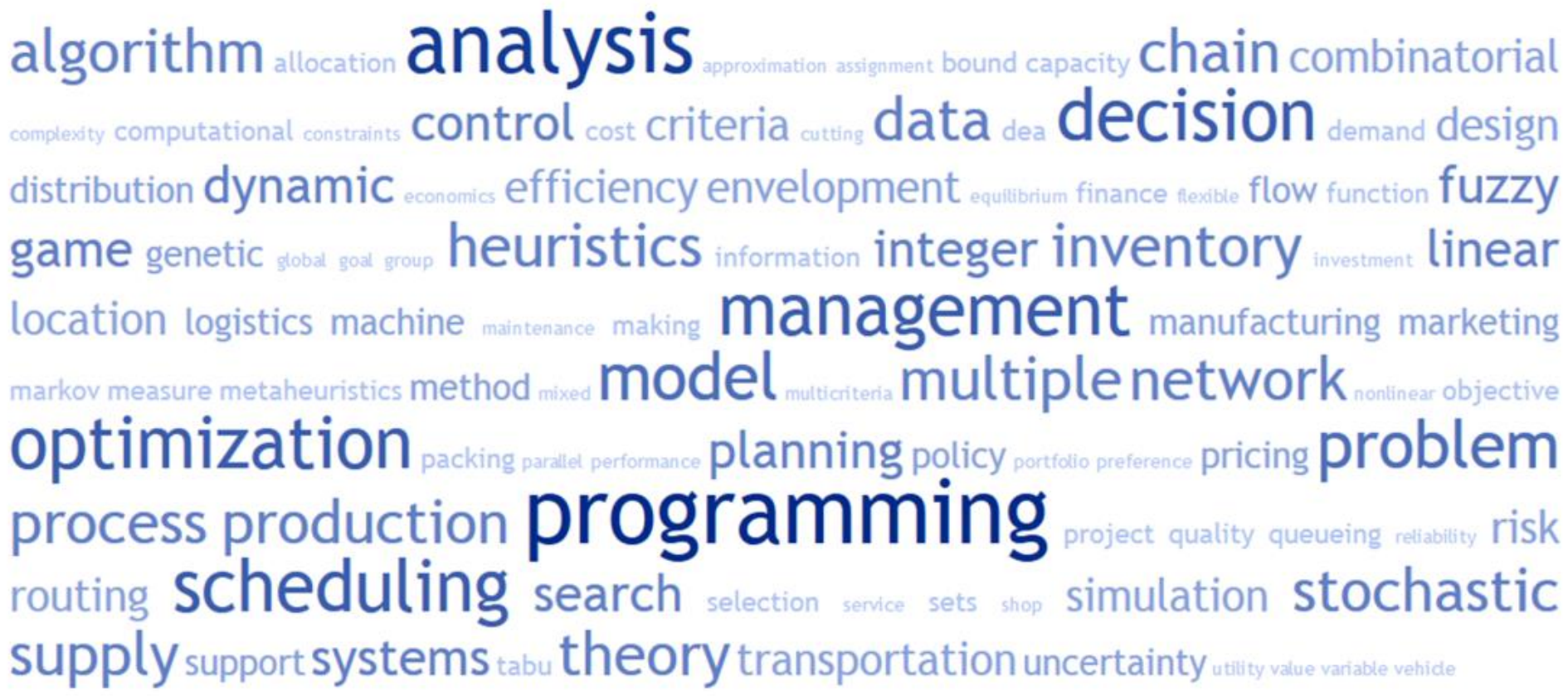

**Fig. 4** Keyword tag crowd showing the top 100 words (TagCrowd.com)

**Table 4** Top 20 most cited papers (articles and reviews) published in journals under the SCI category of OR/MS during 2001–2012

| *Author* | *Year* | *Title* | *Journal* | *# Pages* | *# Ref.* | *# Cites* |
|---|---|---|---|---|---|---|
| Wachter, A; Biegler, LT | 2006 | On the implementation of an interior-point filter line-search algorithm for large-scale nonlinear programming | Math Program | 33 | 29 | 456 |
| Hansen, P; Mladenovic, N | 2001 | Variable neighborhood search: Principles and applications | Eur J Oper Res | 19 | 66 | 438 |
| Dolan, ED; More, JJ | 2002 | Benchmarking optimization software with performance profiles | Math Program | 13 | 18 | 419 |
| Josang, A; Ismail, R; Boyd, C | 2007 | A survey of trust and reputation systems for online service provision | Decis Support Syst | 27 | 73 | 418 |
| Xiao, L; Boyd, S | 2004 | Fast linear iterations for distributed averaging | Syst Control Lett | 14 | 43 | 413 |
| Cachon, GP; Lariviere, MA | 2005 | Supply chain coordination with revenue-sharing contracts: Strengths and limitations | Manage Sci | 15 | 40 | 377 |
| He, Y; Wu, M; She, JH; Liu, GP | 2004 | Delay-dependent robust stability criteria for uncertain neutral systems with mixed delays | Syst Control Lett | 9 | 25 | 370 |
| Fridman, E | 2001 | New Lyapunov-Krasovskii functionals for stability of linear retarded and neutral type systems | Syst Control Lett | 11 | 15 | 359 |
| Frohlich, MT; Westbrook, R | 2001 | Arcs of integration: an international study of supply chain strategies | J Oper Manag | 16 | 84 | 355 |
| Dellarocas, C | 2003 | The digitization of word of mouth: Promise and challenges of online feedback mechanisms | Manage Sci | 18 | 67 | 353 |
| Greco, S; Matarazzo, B; Slowinski, R | 2001 | Rough sets theory for multicriteria decision analysis | Eur J Oper Res | 47 | 102 | 346 |
| Krishnan, V; Ulrich, KT | 2001 | Product development decisions: A review of the literature | Manage Sci | 21 | 208 | 332 |
| Cohen, WM; Nelson, RR; Walsh, JP | 2002 | Links and impacts: The influence of public research on industrial R&D | Manage Sci | 23 | 36 | 331 |
| Argote, L; McEvily, B; Reagans, R | 2003 | Managing knowledge in organizations: An integrative framework and review of emerging themes | Manage Sci | 12 | 76 | 328 |
| Parrilo, PA | 2003 | Semidefinite programming relaxations for semialgebraic problems | Math Program | 28 | 52 | 328 |
| Levin, DZ; Cross, R | 2004 | The strength of weak ties you can trust: The mediating role of trust in effective knowledge transfer | Manage Sci | 14 | 79 | 327 |
| Wang, HZ | 2002 | A survey of maintenance policies of deteriorating systems | Eur J Oper Res | 21 | 180 | 324 |
| Bertsimas, D; Sim, M | 2004 | The price of robustness | Oper Res | 19 | 8 | 310 |
| Borgatti, SP; Cross, R | 2003 | A relational view of information seeking and learning in social networks | Manage Sci | 14 | 92 | 290 |

**Table 5** Top 20 most productive authors from articles published in journals under the SCI category of OR/MS during 2001–2012

| *Author* | *TP (rank)* | *h-index (rank)* | *Country* |
|---|---|---|---|
| Cheng, TCE | 181 (1) | 25 (1) | China |
| Laporte, G | 166 (2) | 23 (2) | Canada |
| Wang, L | 104 (3) | 17 (7) | China |
| Wang, SY | 99 (4) | 19 (3) | China |
| Levitin, G | 95 (5) | 15 (11) | Israel |
| Yang, XQ | 94 (6) | 19 (3) | China |
| Chan, FTS | 94 (6) | 17 (7) | China |
| Sherali, HD | 94 (6) | 15 (11) | USA |
| Wang, Y | 91 (9) | 13 (17) | China |
| Yao, JC | 90 (10) | 19 (3) | Taiwan |
| Tiwari, MK | 90 (10) | 15 (11) | India |
| Berman, O | 88 (12) | 13 (17) | Canada |
| Kim, J | 88 (12) | 12 (19) | South Korea |
| Pardalos, PM | 84 (14) | 18 (6) | USA |
| Yang, H | 84 (14) | 17 (7) | China |
| Kumar, S | 84 (14) | 14 (14) | USA |
| Li, J | 83 (17) | 11 (20) | China |
| Lai, KK | 80 (18) | 14 (14) | China |
| Teo, KL | 76 (19) | 17 (7) | China |
| Xie, M | 72 (20) | 14 (14) | Singapore |

*TP* total publications